\documentclass{amsart}
\newtheorem{theorem}{Theorem}[section]

\theoremstyle{definition}
\newtheorem{definition}[theorem]{Definition}

\theoremstyle{remark}
\newtheorem{remark}[theorem]{Remark}
\numberwithin{equation}{section}
\newcommand{\abs}[1]{\lvert#1\rvert}

\newcommand{\N}{\mathcal{N}}
\newcommand{\HI}{\mathfrak{H}}

\newcommand{\C}{\mathbb{C}}
\newcommand{\D}{\mathfrak{D}}
\newcommand{\HQ}{\mathbb{H}}

\begin{document}
\title{Coherent states associated to the wavefunctions and the spectrum of the isotonic oscillator}
\author{K. Thirulogasanthar}
\address{Department of Mathematics and Statistics, Concordia University,  
   7141 Sherbrooke Street West, Montreal, Quebec H4B 1R6, Canada }
\email{santhar@vax2.concordia.ca}




\author{Nasser Saad}
\address{Department of Mathematics and Statistics, University of Prince Edward Island, 550 University Avenue, Charlottetown, PEI, C1A 4P3, Canada.}
\email{nsaad@upei.ca}


\date{\today}



\begin{abstract} 

Classes of coherent states are presented by replacing the labeling parameter $z$ of Klauder-Perelomov type coherent states by confluent hypergeometric functions with specific parameters. Temporally stable coherent states for the isotonic oscillator Hamiltonian are presented and these states are identified as a particular case of the so-called Mittag-Leffler coherent states. 
\end{abstract}

\maketitle

\pagestyle{myheadings}

\markboth{K.Thirulogasanthar and Nasser Saad}{CS for the isotonic oscillator}



\section{Introduction}

Hilbert spaces are the skeleton of the mathematical aside of quantum theories. Coherent states (CS for short) are defined as a specific overcomplete family of vectors in the Hilbert space of the problem to describe quantum phenomenon. In particular, CS are mathematical tools which provide a close connection between classical and quantum formalisms. There are several ways to define a set of CS \cite{Ali}-\cite{Pere}. In this article we follow the generalizations of the cannonical CS.
Let $\{\phi_m\}_{m=0}^{\infty}$ be an orthonormal basis of a separable Hilbert space $\HI$. For $z\in\C$, the complex plane, the well-known canonical coherent states are defined as
\begin{equation}
\mid z\rangle=e^{-\frac{r^2}{2}}\sum_{m=0}^{\infty}\frac{z^m}{\sqrt{m!}}\phi_m\in\HI,
\end{equation}
where $z=re^{i\theta}$. This definition has been generalized as follows: Let $\D$ be an open subset of $\C$. For $z\in\D$ define
\begin{equation}\label{can-cs}
\mid z\rangle=\N(|z|)^{-\frac{1}{2}}\sum_{m=0}^{\infty}\frac{z^m}{\sqrt{\rho(m)}}\phi_m\in\HI
\end{equation}
where $\{\rho(m)\}_{m=0}^{\infty}$ is a positive sequence of real numbers and $\N(|z|)$ is the normalization constant ensuring that $\langle z\mid z\rangle=1.$ In addition, if the set $\{\mid z\rangle: z\in \D\}$ satisfies 
\begin{equation}\label{can-res}
\int_{\D}\mid z\rangle\langle z\mid d\mu=I_{\HI},
\end{equation}
where $d\mu$ is an appropriately chosen measure on $\D$, then $\{\mid z\rangle: z\in \D\}$ is a set of coherent states on $\D$. A further generalization can be motivated as follows: Let $(X,\mu)$ be a measure space and $\HQ\subseteq L^2(X,\mu)$ be a closed subspace. Let $\{\Phi_m\}_{m=0}^{\text{dim}{(\HQ)}}$ be an orthonormal basis of $\HQ$ satisfying, for arbitrary $x\in X$,
$$\sum_{m=0}^{\text{dim}{(\HQ)}}\abs{\Phi_m(x)}^2<\infty,$$
where $\text{dim}{(\HQ)}$ denotes the dimension of $\HQ$. Let $\HI$ be another Hilbert space with $\text{dim}(\HQ)=\text{dim}(\HI)$ and $\{\phi_m\}_{m=0}^{\text{dim}{(\HI)}}$ be an orthonormal basis of $\HI$. Define
\begin{equation}\label{rep}
K(x,y)=\sum_{m=0}^{\text{dim}{(\HQ)}}\overline{\Phi_m(x)}\Phi_{m}(y).
\end{equation}
Then $K(x,y)$ is a reproducing kernel and $\HQ$ is the corresponding reproducing kernel Hilbert space. For $x\in X$, define
\begin{equation}\label{rep-cs}
\mid x\rangle=K(x,x)^{-\frac{1}{2}}\sum_{m=0}^{\text{dim}{(\HQ)}}\Phi_m(x)\phi_m.
\end{equation}
Then, it is straightforward to show that
$$\langle x\mid x\rangle=K(x,x)^{-1}\sum_{m=0}^{\text{dim}{(\HQ)}}\overline{\Phi_m(x)}\Phi_m(x)=1$$
and
$$\mathcal{W}:\HI\longrightarrow\HQ\;\;\;\text{with}\;\;\;\mathcal{W}\phi(x)=K(x,x)^{\frac{1}{2}}\langle x\mid \phi\rangle$$
is an isometry. Thus, for $\phi,\psi\in\HI$, we have
\begin{eqnarray*}
\langle\phi\mid\psi\rangle_{\HI}=\langle\mathcal{W}\phi\mid\mathcal{W}\psi\rangle_{\HQ}
=\int_{\Omega}\langle\phi\mid x\rangle\langle x\mid\psi\rangle K(x,x)d\mu(x),
\end{eqnarray*}
and thereby we have a resolution of the identity
\begin{equation}\label{res-rep}
\int_{X}\mid x\rangle\langle x\mid K(x,x)d\mu(x)=I_{\HI},
\end{equation}
where $K(x,x)$ appears as a positive weight function. Thus the set of states $\{\mid x\rangle: x\in X\}$ forms a set of CS. In the case where $\{\Phi_m\}_{m=0}^{\text{dim}{(\HQ)}}$ is an orthogonal basis of $\HQ$, one can define $\rho(m)=\|\Phi_m\|^2;\;m=0,...,\text{dim}{(\HQ)},$ and get an orthonormal basis $\left\{{\Phi_m}/{\sqrt{\rho(m)}}\right\}_{m=0}^{\text{dim}{(\HQ)}}$ of $\HQ$ and set
\begin{equation}\label{rep-cs-2}
\mid x\rangle=K(x,x)^{-\frac{1}{2}}\sum_{m=0}^{\text{dim}(\HI)}\frac{\Phi_m(x)}{\sqrt{\rho(m)}}\phi_m\in\HI,
\end{equation}
which is a generalization of (\ref{can-cs}). The above discussion provoke the following definition.
\begin{definition}\label{D1}
Let $\mathfrak H$ be a separable Hilbert space with an orthonormal basis $\{\phi_{m}\}_{m=0}^{\infty}$. Let $\D$ be an open subset of $\mathbb C$ and let
$$\Phi_m: \D\longrightarrow \C,\;\;\;m=0,1,2,\dots,$$
a sequence of complex functions. Define 
\begin{equation}\label{ma-cs}
\mid z\rangle=\mathcal N(|z|)^{-\frac{1}{2}}\sum_{m=0}^{\infty}\frac{\Phi_{m}(z)}{\sqrt{\rho(m)}}\phi_{m}\in\mathfrak{H};\;\;z\in\D,
\end{equation}
where $\mathcal N(|z|)$ is a normalization factor and $\{\rho(m)\}_{m=0}^{\infty}$ is a sequence of nonzero positive real numbers. The set of vectors in (\ref{ma-cs}) 
are said to form a set of CS if
\begin{enumerate}
\item[(a)]
For each $z\in\D$, the states $\mid z\rangle$ are normalized, that is, $\langle z\mid z\rangle=1;$
\item[(b)]
The states $\{\mid z\rangle\;:\;z\in\D\}$ satisfy a resolution of the identity:
\begin{equation}
\int_{\mathfrak D}\mid z\rangle \langle z\mid d\mu=I\label{2},
\end{equation}
\end{enumerate}
where $d\mu$ is an appropriately chosen measure and $I$ is the identity operator on $\mathfrak H$.
\end{definition}

For $\Phi_m(z)=z^m$, $z\in\C$, the states (\ref{ma-cs}) have been studied extensively in quantum theories for various $\rho(m)$ \cite{Ali,Ks,Kp}. Recently, Gazeau and Klauder introduced a class of CS for Hamiltonians with discrete and continuous spectrum by taking $$\Phi_m(J,\alpha)={J}^{m/2}e^{ie_m\alpha}$$ and $\rho(m)=e_1e_2...e_m$, where $e_m$'s are the spectrum of the Hamiltonian arranged in a suitable manner \cite{Gk}. In their definition, next to the conditions (a) and (b), the following are added:
\begin{enumerate} 
\item[(c)] Temporal stability;
\item[(d)] Action identity.
\end{enumerate}
Following the method proposed by Gazeau and Klauder, several new classes of Gazeau-Klauder type CS were investigated \cite{Ahmed}-\cite{Bor}. In \cite{Ta} vector coherent states were constructed by taking $\Phi_m(Z)=Z^m$ where $Z$ is an $n\times n$ matrix valued function.

Motivated by the recent interest, in Section 2 we present a class of orthonormal basis emerge from the exact solutions of the isotonic oscillator \cite{LL}. In Section 3, we present two new classes of CS labeled by confluent hypergeometric functions ${}_1F_1$ with specific parameters. In Section 4, we present temporally stable CS for the isotonic oscillator. By temporally stable CS we mean a class of CS satisfying conditions (a)-(c). We also identify these states as an analogue of the Mittag-Leffler CS \cite{SPS}.
\section{Isotonic oscillator wavefunctions as orthonormal basis}
\label{sec2}
\noindent A few of the quantum Hamiltonians known to have exact solutions. An interesting model of a solvable class is the isotonic oscillator Hamiltonian \cite{LL}
\begin{equation}\label{h1}
H=-\frac{d^2}{dx^2}+x^2+\frac{A}{x^2}\;\;\;(A\geq 0),
\end{equation}
acting in the Hilbert space $L^{2}(\mathbb R^{+},dx)$ and the eigenfunctions $\psi\in L^2(0,\infty)$ satisfy the Dirichlet boundary condition; $\psi(0)=0$. The Hamiltonian is the generalization of the harmonic oscillator Hamiltonian in 3-dimension where the generalization lies in the parameter $A$ ranging over [$0,\infty)$ instead of  the angular momentum quantum numbers $l=0,1,2,\dots$.
An interesting feature of this Hamiltonian is that it has singularity coincides with that of the spiked harmonic oscillator Hamiltonian,
\begin{equation}
H_\alpha=-\frac{d^2}{dx^2}+x^2+\frac{\lambda}{x^\alpha},\;\;\;\;\;(\alpha>0,\lambda>0).
\end{equation}
It has been proved that the basis constructed from the exact solutions of the isotonic oscillator Hamiltonian $H$ form a more effective starting point for a perturbative variational treatment \cite{RNA} of the singular Hamiltonian $H_\alpha$ than the basis of the ordinary harmonic oscillator Hamiltonian $(A=0)$. It is known that the Hamiltonian $H$ admits exact solutions \cite{LL,RNA},
\begin{equation}\label{h2}
\psi_{m}^{\gamma}(x)=(-1)^{m}\sqrt{\frac{2(\gamma)_m}{m!\Gamma(\gamma)}}x^{\gamma-\frac{1}{2}}e^{-\frac{1}{2}x^2}{}_1F_1(-m;\gamma;x^{2}),\;\;\;m=0,1,2,..
\end{equation}
where $\gamma=1+\frac{1}{2}\sqrt{1+4A}$; and the exact eigenvalues are given by

\begin{equation}\label{h3}
e_{m}=2(2m+\gamma),\;\;\;m=0,1,2,...
\end{equation}
It was also shown that (see \cite{RNA}, especially section 3)
\begin{equation}\label{h4}
\int_{0}^{\infty}\psi_{m}^{\gamma}(x)\psi_{n}^{\gamma}(x)dx=\delta_{nm},
\end{equation}
and the collection of vectors $\{\mid \psi_{m}^{\gamma}\rangle\}_{m=0}^{\infty}$ form an orthonormal basis for the Hilbert space $\widehat{\mathfrak{H}}=L^{2}(\mathbb R^{+},dx).$ 
\section{CS with confluent hypergeometric functions}
\noindent In this section we present two classes of CS in the state Hilbert space of the isotonic oscillator using Definition \ref{D1}. The first class is built with 
$$\Phi_m(x,\theta)=e^{im\theta}{}_1F_{1}(-m;\gamma;x^2)\;\;\text{and}\;\;\rho(m)=\frac{m!(\frac{\gamma}{2}+m)}{(\gamma)_m}$$
and the second is built with 
$$\Phi_m(x,\theta)=\sqrt{{}_1F_{1}(-m;\gamma+1;x)}e^{im\theta}\;\;\text{and}\;\;\rho(m)=\frac{\gamma}{\gamma+m}.$$
At the end of this section we will discuss the convenience of these choices together with a general discussion.
\subsection{Class-I} For $\gamma>2$, let us denote

$${}_1F_{1}^{\theta}(-m;\gamma;x^2)=e^{im\theta}{}_1F_{1}(-m;\gamma;x^2)$$
and consider the following set of states
\begin{equation}\label{s31}
\mid x,\theta,\gamma\rangle=\mathcal N(x,\gamma)^{-\frac{1}{2}}\sum_{m=0}^{\infty}\sqrt{\frac{(\gamma)_m}{m!(\frac{\gamma}{2}+m)}}{}_1F_{1}^{\theta}(-m;\gamma;x^2)\mid\psi_{m}^{\gamma}\rangle.
\end{equation}
We show that the states (\ref{s31}) form a set of CS in the sense of Definition \ref{D1}. In order to identify the normalization factor, we consider
$$
1=\langle x,\theta,\gamma\mid x,\theta,\gamma\rangle=\mathcal N(x,\gamma)^{-1}\sum_{m=0}^{\infty}\frac{(\gamma)_m}{m!(\frac{\gamma}{2}+m)}[{}_1F_{1}(-m;\gamma;x^2)]^{2}.$$
Thereby, for $0<x<\infty,$ we have
\begin{equation}\label{s32}
{\mathcal N}(x,\gamma)=\sum_{m=0}^{\infty}\frac{(\gamma)_m}{m!(\frac{\gamma}{2}+m)}[{}_1F_{1}(-m;\gamma;x^2)]^{2}
= \frac{\Gamma(\gamma)e^{x^2}}{x^{2(\gamma-1)}}K_{\frac{\gamma-1}{2}}\left(\frac{x^2}{2}\right)I_{\frac{\gamma-1}{2}}\left(\frac{x^2}{2}\right),
\end{equation}
where $I_{\nu}(x)$ and $K_{\nu}(x)$ are modified Bessel functions of the first and second kind respectively. The proof of this summation formula is obtained by means of constructing the Green's function for the Hamiltonian $H$ and then applying a Mercer type theorem that arises in connection with integral equations to obtain the uniform convengence of this sum. A detail proof of this identity and related series can be found in \cite{Attila}. For the resolution of the identity, let us assume that the measure takes the form
$$
d\mu(x,\theta)=\frac{\mathcal N(x,\gamma)}{2\pi}\lambda(x)dxd\theta,
$$
where $\theta\in[0,2\pi)$ and $\lambda(x)$ is an auxiliary density to be determined shortly. Since
$$\int_{0}^{2\pi}e^{i(m-n)\theta}d\theta=\left\{\begin{array}{ccc}
0&\text{if}&m\not=n\\
1&\text{if}&m=n
\end{array}\right.$$
we have, if there is a density $\lambda(x)$ to satisfy
\begin{equation}\label{s33}
\int_{0}^{\infty}[{}_1F_{1}(-m;\gamma;x^2)]^{2}\lambda(x)dx=\frac{m!(\frac{\gamma}{2}+m)}{(\gamma)_m},
\end{equation}
that
\begin{eqnarray*}
&&\int_{0}^{\infty}\int_{0}^{2\pi}\mid x,\theta,\gamma\rangle\langle x,\theta,\gamma\mid d\mu(x,\theta)\\
&=&\sum_{m=0}^{\infty}\bigg[\frac{(\gamma)_m}{m!(\frac{\gamma}{2}+m)}\int_{0}^{\infty}[{}_1F_{1}(-m;\gamma;x^2)]^{2}\lambda(x)dx\bigg]\mid\psi_{m}^{\gamma}\rangle\langle\psi_{m}^{\gamma}\mid\\
&=&\sum_{m=0}^{\infty}\mid\psi_{m}^{\gamma}\rangle\langle\psi_{m}^{\gamma}\mid=I_{\widehat{\mathfrak{H}}}.
\end{eqnarray*}
For $2\gamma-\alpha>0$ we have (see \cite{Nh}, especially Eq.(4.5) or \cite{LL} formula f.7)
\begin{eqnarray*}
&&\int_{0}^{\infty}x^{2\gamma-\alpha-1}e^{-x^2}[{}_1F_1(-n;\gamma;x^2)]^2dx\\
&&\hspace{2cm}=\frac{\left(\frac{\alpha}{2}\right)_{n}\Gamma(\gamma-\frac{\alpha}{2})}{2(\gamma)_{n}}{}_3F_2(-m,\gamma-\frac{\alpha}{2},1-\frac{\alpha}{2};\gamma,1-\frac{\alpha}{2}-n;1),
\end{eqnarray*}
which yields in tha case of $\alpha=4$ that
\begin{equation}\label{s35}
\int_{0}^{\infty}x^{2\gamma-5}e^{-x^2}[{}_1F_{1}(-m;\gamma;x^2)]^2dx=\frac{m!\Gamma(\gamma-2)}{\gamma(\gamma)_m}\left[\frac{\gamma}{2}+m\right].
\end{equation}
Consequently, the density $\lambda(x)$ that satisfies (\ref{s33}) is given by
$$\lambda(x)=\frac{\Gamma(\gamma-2)}{\gamma}x^{2\gamma-5}e^{-x^2}.$$
Thus the set of states (\ref{s31}) forms a set of CS.
\subsection{Class-II}

For $\gamma>1$, consider the following set of states in the Hilbert space $\widehat{\mathfrak{H}}.$
\begin{equation}\label{s36}
\mid x,\theta,\gamma\rangle=\mathcal N(x,\gamma)^{-\frac{1}{2}}\sum_{m=0}^{\infty}\sqrt{\frac{(\gamma+m){}_1F_{1}(-m;\gamma+1;x)}{\gamma}}e^{im\theta}\mid\psi_{m}^{\gamma}\rangle
\end{equation}
where the normalization factor is given by 
\begin{equation}\label{s37}
{\mathcal N}(x,\gamma)=\sum_{m=0}^{\infty}\frac{(\gamma+m){}_1F_{1}(-m;\gamma+1;x)}{\gamma}=\frac{\gamma-1}{x}+\frac{\gamma-1}{x^2};\;\;\;\;x>0
\end{equation}
by means of the Buchholz's identity \cite{buc}
\begin{equation}\label{s38}
\sum_{n=0}^{\infty}\frac{(-\nu)_n\Gamma(\gamma+\nu+1)}{n!\Gamma(\gamma+1)}{}_1F_1(-n;\gamma+1;y)=y^\nu,\quad\quad \gamma+\nu>-1,
\end{equation}
for $\nu=-1$ and $\nu=-2$ respectively. Thus the normalization factor is defined for all $x>0$ and it is positive for all $\gamma>1$.  
For $x\in[0,\infty)$ and $\theta\in[0,2\pi)$, with a positive density function $\lambda(x)$, assume
$$d\nu(x,\theta)=\frac{\mathcal N(x,\gamma)}{2\pi}\lambda(x)dxd\theta.$$
A resolution of the identity can be obtained as follows:
\begin{eqnarray*}
&&\int_{0}^{\infty}\int_{0}^{2\pi}\mid x,\theta,\gamma\rangle\langle x,\theta,\gamma\mid d\nu(x,\theta)\\
&&\hspace{1cm}=\sum_{m=0}^{\infty}\bigg[\frac{(\gamma+m)}{\gamma}\int_{0}^{\infty}{}_1F_{1}(-m;\gamma+1;x)\lambda(x)dx\bigg]\mid\psi_{m}^{\gamma}\rangle\langle\psi_{m}^{\gamma}\mid\\
&&\hspace{1cm}=\sum_{m=0}^{\infty}\mid\psi_{m}^{\gamma}\rangle\langle\psi_{m}^{\gamma}\mid=I_{\widehat{\mathfrak{H}}}
\end{eqnarray*}
if there is a density $\lambda(x)$ such that
\begin{equation}\label{s39}
\int_{0}^{\infty}{}_1F_{1}(-m;\gamma+1;x)\lambda(x)dx=\frac{\gamma}{(\gamma+m)}.
\end{equation}
Since (see Eq.(2.4) of \cite{Nh})
\begin{equation}\label{s310}
\int_{0}^{\infty}x^{d-1}e^{-x}{}_1F_1(a;b;x)dx=\Gamma(d){}_2F_{1}(a,d;b;1)
\end{equation}
if we take $a=-m,\;d=1$, we have using Chu-Vandermonde's theorem that
\begin{equation}\label{s311}
\int_{0}^{\infty}e^{-x}{}_1F_{1}(-m;b;x)dx={}_2F_{1}(-m,1;b;1)=\frac{(b-1)_m}{(b)_m}=\frac{(b-1)}{(b-1+m)}
\end{equation}
which justify (\ref{s39}) with $\lambda(x)=e^{-x}$. Thus the states (\ref{s36}) form a set of CS.

\begin{remark}
$\bullet$~In a general setting, the normalization condition, $\langle z\mid z\rangle=1$ of Definition \ref{D1} requires the following:
\begin{equation}\label{s312}
\N(|z|)=\sum_{m=0}^{\infty}\frac{\abs{\Phi_m(z)}^2}{\rho(m)}<\infty.
\end{equation}
In the same definition if we set a measure, $d\mu(z)=\N(|z|)d\nu(z)$ on $\D$, a resolution of the identity can be guaranteed by the orthogonality relation
\begin{equation}\label{s313}
\int_{\D}\Phi_m(z)\Phi_n(z)d\nu(z)=\delta_{mn}\rho(m).
\end{equation}
When we choose $\Phi_m,\rho(m)$ and $\D$ to define a set of CS of type (\ref{ma-cs}) our choices need to satisfy (\ref{s312}) and (\ref{s313}) simultaneously. In defining the CS, (\ref{s31}) and (\ref{s36}), the term $e^{im\theta}$ is introduced to get the orthogonality among $\Phi_m(x,\theta)$. In the view of (\ref{s312}) and (\ref{s313}), the choices in the class of CS (\ref{s31}) are depicted by the convergence of the series (\ref{s32}) and the existence of a positive density $\lambda(x)$ to satisfy the identity (\ref{s33}). Similarly, the choices of CS (\ref{s36}) are predicted by the identities (\ref{s37}) and (\ref{s310}). The condition (\ref{s313}) reminds us of the classical orthogonal polynomials, however the $\rho(m)$ of the usual orthogonality relations of classical polynomials do not satisfy (\ref{s312}). It may be interesting to look for a way to use those orthogonality relations to define CS labeled by classical polynomials  without introducing the additional term $e^{im\theta}$.\\
$\bullet$~ For the CS of type (\ref{can-cs}) one can naturally associate a generalized oscillator algebra with generalized annihilation ($\mathbf{a}$), creation ($\mathbf{a}^{\dagger}$) and  number ($\mathbf{n}$) operators \cite{Ali,Bor,Ta,Th}. The generalized oscillator algebra is the Lie algebra generated by the set of operators $\{\mathbf{a},\mathbf{a}^{\dagger},\mathbf{n}\}$. In this case, the set of CS (\ref{can-cs}) satisfies
 \begin{equation}\label{d1al}
\mathbf{a} \mid z \rangle = z \mid z \rangle.
\end{equation}
Following the same method, for the CS of Definition \ref{D1} a class of operators cannot be defined to satisfy a relation similar to (\ref{d1al}) unless $\Phi_m(z)=[f(z)]^m$ for some function $f(z)$. This point is discussed in detail together with examples in \cite{Th}.\\
$\bullet$~For the states (\ref{s31}) and (\ref{s36}), through the resolution of the identity, a reproducing kernel and a reproducing kernel Hilbert space can be associated using the standard procedure \cite{Ali}. Along these lines, for the CS of Definition \ref{D1} a detail explanation is also given in \cite{Th}, which can be adapted to the states (\ref{s31}) and (\ref{s35}) in a straightforward manner.\\
$\bullet$~The states (\ref{s31}) and (\ref{s36}) are defined as vectors in the state Hilbert space of the isotonic oscillator. In the case where these states describe a quantum system, by the usual quantum mechanical convention, the probability of finding the state $\mid\psi_{m}^{\gamma}\rangle$ in some normalized state $\mid z\rangle$ of the state Hilbert space is given by
$$P(m,z,\gamma)=\abs{\langle\psi_{m}^{\gamma}\mid z\rangle}^2.$$
Thus for the states (\ref{s31}) it is given by
$$P(m,x,\gamma,\theta)=\frac{(\gamma)_m}{\N(x,\gamma)m!(\frac{\gamma}{2}+m)}[{}_1F_{1}(-m;\gamma;x^2)]^2$$
where as for the states (\ref{s36}) it is given by
$$P(m,x,\gamma,\theta)=\frac{(\gamma+m)\abs{{}_1F_{1}(-m;\gamma+1;x)}}{\gamma\N(x,\gamma)}.$$
For a state $\mid\psi\rangle$ of the state Hilbert space the average energy of the system is given by
$E=\langle\psi\mid H\mid\psi\rangle.$
Thereby, for the CS of (\ref{s36}) we have
$$E=\N(x,\gamma)^{-1}\sum_{m=0}^{\infty}\frac{2(\gamma+m)(\gamma+2m)}{\gamma}{}_1F_{1}(-m;\gamma+1;x^2)$$
where the convergence of the infinite sum follows by means of the Buchholz's identity (\ref{s38}), and which yields
$$E=\frac{2}{x^6\N(x,\gamma)}(x^2-x+2)(x^2+x+2)(\gamma-1)(\gamma-2)$$
here $\N(x,\gamma)$ is given by (\ref{s47}).\\

The sets of CS (\ref{s36}) and (\ref{s31}) are derived with the orthonormal basis of the state Hilbert space of the Hamiltonian $H$, which can be replaced by any other separable Hilbert space, for example if one replaces it with the Fock space of the Harmonic oscillator Hamiltonian, say $H_1$, then in the expressions (\ref{s36}) and (\ref{s31}) the basis $\mid \psi_{m}^{\gamma}\rangle$ has to be replaced by the Fock space basis $\mid m\rangle$ and the parameter $\gamma$ not necessarily be associated to the Hamiltonian $H$. In this case,  the average energy, $E_1=\langle x,\gamma\mid H_1\mid x,\gamma \rangle$, can also be calculated as long as it is physically valid.\\

For the states of Definition \ref{D1}, in an abstract way, one can also define a Hamiltonian with the orthonormal basis of the Hilbert space. For this, let
$$y_m=\frac{\rho(m)}{\rho(m-1)},\;\;\;\forall m\geq 1$$
and assume $y_{0}:=0$. Then $\rho(m)=y_1\dots y_m=y_m!$. Define a new Hamiltonian by
\begin{equation}\label{ha-1}
H_2=\sum_{m=0}^{\infty}y_m\mid\phi_{m}\rangle\langle\phi_{m}\mid.
\end{equation}
Then we have $H_2\mid\phi_{m}\rangle=y_m\mid\phi_{m}\rangle$ for $m\geq 0$. If we calculate the average energy of the CS (\ref{ma-cs}) with the Hamiltonian $H_2$, we get,
\begin{eqnarray}
E_2=\langle z\mid H_2\mid z\rangle&=&\frac{1}{\N(|z|)}\sum_{m=0}^{\infty}\frac{\abs{\Phi_{m+1}(z)}^2}{y_m!}\nonumber
\end{eqnarray}
Following the above procedures the CS of (\ref{s31}) can be associated to different Hamiltonians and the average energy can also be calculated, however, in this case, determining the finiteness of the energy may be troublesome.\\

\end{remark}
\section{Coherent states for isotonic oscillator}
\label{sec3} 
\noindent Let $H$ be a Hamiltonian with a bounded below discrete spectrum $\{e_m\}$ and it has been adjusted so that $H\geq 0$. Further assume that the eigenvalues $e_m$ are non-degenerate and arranged in increasing order, $e_0<e_1<...$. For such a Hamiltonian, a class of CS was suggested by Gazeau and Klauder \cite{Gk}, the so-called {\em Gazeau-Klauder coherent states}, as
\begin{equation}\label{s41}
\mid J,\alpha\rangle=\mathcal N(J)^{-1}\sum_{m=0}^{\infty}\frac{J^{{m/2}}}{\sqrt{\rho(m)}}e^{-ie_n\alpha}\mid m\rangle,
\end{equation}
where $J\geq 0$, $-\infty<\alpha<\infty$, $\rho(m)=e_1e_2e_3...e_m=e_m!$ and $\mid m\rangle$ are the eigenfunctions of $H$.\\
For the isotonic oscillator Hamiltonian (\ref{h1}), the eigenvalues (\ref{h3}) satisfy 
$$2\gamma=e_0<e_1<e_2<e_3<....$$
therefore
$$\rho(m)=4^{m}(\frac{\gamma}{2}+1)_m.$$
Consequently, the states (\ref{s41}) takes the form
\begin{equation}\label{s42}
\mid J,\alpha\rangle=\mathcal N(J)^{-1}\sum_{m=0}^{\infty}\frac{J^{{m}/{2}}e^{-2i(2m+\gamma)\alpha}}{\sqrt{4^m(1+\frac{\gamma}{2})_m}}\mid\psi_{m}^{\gamma}\rangle.
\end{equation}
The normalization factor 
\begin{equation}\label{s43}
{\mathcal N}(J)^2=\sum_{m=0}^{\infty}\frac{J^{m}}{4^m(\frac{\gamma}{2}+1)_m}={}_1F_1(1;\gamma+1;\frac{J}{4}).
\end{equation}
Let the integral on $\alpha$ be defined by
$$\int...d\alpha=\lim_{\delta\rightarrow\infty}\frac{1}{2\delta}\int_{-\delta}^{\delta}...d\alpha.$$
Notice that,
$$\int e^{-i\alpha(e_m-e_l)}d\alpha=\lim_{\delta\rightarrow\infty}\frac{1}{2\delta}\int_{-\delta}^{\delta} e^{-i\alpha(e_m-e_l)}d\alpha=\left\{\begin{array}{ccc}
0&\text{if}&m\not=l\\
1&\text{if}&m=l
\end{array}\right.$$
For a resolution of the identity we notice that the radius of convergence
$$R=\lim_{m\longrightarrow\infty}\sqrt[m]{\rho(m)}=\infty.$$
For $J\in[0,\infty)$ and $\alpha\in(-\infty,\infty)$, let 
$$d\mu(J,\alpha)=\mathcal N(J)\lambda(J)dJd\alpha,$$
consequently, the resolution of identity follows by mean of 
\begin{equation}\label{s44}
\int_{0}^{\infty}\int\mid J,\alpha\rangle\langle J,\alpha\mid \lambda(J)d\alpha dJ
=\sum_{m=0}^{\infty}\mid\psi_{m}^{\gamma}\rangle\langle\psi_{m}^{\gamma}\mid=I_{\widehat{\mathfrak{H}}}
\end{equation}
provided that $\lambda(J)$ satisfies
\begin{equation}\label{s45}
\int_{0}^{\infty}J^{m}\lambda(J)dJ=4^m(\frac{\gamma}{2}+1)_m.
\end{equation}
From the Mellin transform \cite{mag},
\begin{equation}\label{s46}
\int_{0}^{\infty}e^{-ax}x^{s-1}dx=a^{-s}\Gamma(s)
\end{equation}
it is evident that the weight
$$\lambda(J)=\frac{{J}^{-\gamma/2}e^{-J/4}}{2^{\gamma+2}\Gamma(1+\frac{\gamma}{2})}$$
satisfies (\ref{s45}). Now for the temporal stability, since
$$H\mid\psi_{m}^{\gamma}\rangle=e_{m}\mid\psi_{m}^{\gamma}\rangle,\;\;\;\;m=0,1,2,...$$
and
$$e^{-ie_n\alpha}e^{-iHt}\mid\psi_{m}^{\gamma}\rangle=e^{-ie_n\alpha}e^{-ie_nt}\mid\psi_{m}^{\gamma}\rangle=e^{-ie_n(\alpha+t)}\mid\psi_{m}^{\gamma}\rangle$$
we have
\begin{equation}\label{s47}
e^{-iHt}\mid J,\alpha\rangle=\mid J,\alpha+t\rangle.
\end{equation}
Thus the states $\mid J,\alpha\rangle$ are temporally stable. Since, for the isotonic oscillator, $e_0=2\gamma\not=0$ we cannot get the action identity, i.e., $\langle J,\alpha\mid H\mid J,\alpha\rangle\not=J$.
In this respect the states in (\ref{s42}) form a set of temporally stable CS for the isotonic oscillator but not a set of Gazeau-Klauder CS. One can easily calculate the overlap of two states:
\begin{eqnarray*}
\langle J',\alpha'\mid J,\alpha\rangle&=&\frac{1}{\mathcal N(J')\mathcal N(J)}\sum_{m=0}^{\infty}\frac{(J'J)^{\frac{m}{2}}}{4^m(\frac{\gamma}{2}+1)_m}e^{-2i(2m+\gamma)(\alpha-\alpha')}\\&=& \frac{e^{-2i\gamma(\alpha-\alpha^\prime)}}{{\mathcal N}(J^\prime){\mathcal N}(J)}
{}_1F_1(1;\frac{\gamma}{2}+1;e^{-4i\gamma(\alpha-\alpha^\prime)}\frac{\sqrt{JJ'}}{4}).
\end{eqnarray*}
In particular, if $\alpha=\alpha'$ we get
\begin{eqnarray*}
\langle J',\alpha\mid J,\alpha\rangle&=&\frac{1}{\mathcal N(J')\mathcal N(J)}\sum_{m=0}^{\infty}\frac{\Gamma(1+\frac{\gamma}{2})(J'J)^{\frac{m}{2}}}{4^m\Gamma(m+1+\frac{\gamma}{2})}\\&=&\frac{{}_1F_1(1;\frac{\gamma}{2}+1;\frac{\sqrt{JJ'}}{4})}{\sqrt{{}_1F_1(1;\frac{\gamma}{2}+1;\frac{J}{4}){}_1F_1(1;\frac{\gamma}{2}+1;\frac{J'}{4})}}.
\end{eqnarray*}
\begin{remark}
$\bullet$~Let us shift the spectrum $e_m$ as $\epsilon_m=e_m-e_0=4m$ then we have $0=\epsilon_0<\epsilon_1<...$. Instead of taking $\rho(m)=e_1\dots e_m$ in (\ref{s42}) if we take $\rho(m)=\epsilon_1\dots\epsilon_m=4^m\Gamma(m+1)$ we can satisfy the action identity condition, and, thereby, we can  have Gazeau-Klauder CS. In this case, the normalization factor is given by $\N(J)=e^{J/8}$ and a resolution of the identity can be obtained with the measure $d\mu(J,\alpha)=\frac{d\alpha dJ}{4}$ by means of (\ref{s46}). The temporal stability and the action identity can easily be verified. However, this case is simpler and it is, at least computationally, similar to the ordinary harmonic oscillator.\\
$\bullet$ Sixdeniers {\it et al} \cite{SPS}  considered the so-called Mittag-Leffler CS, using the Mittag-Leffler functions
$$E_{a,b}=\sum_{m=0}^{\infty}\frac{x^m}{\Gamma(am+b)};\;\;\;a,b>0,$$
as follows:
\begin{equation}\label{s48}
\mid z,a,b\rangle=\N_{a,b}(\abs{z}^2)^{-\frac{1}{2}}\sum_{m=0}^{\infty}\frac{z^m\sqrt{\Gamma(b)}}{\sqrt{\Gamma(am+b)}}\mid m\rangle.
\end{equation}
The normalization factor is given by
$$\N_{a,b}(\abs{z}^2)=\Gamma(b)E_{a,b}(\abs{z}^2)$$
and a resolution of the identity is obtained as
$$\int\int_{\C}d^2z\mid z,a,b\rangle\widetilde{W}_{a,b}(\abs{z}^2)\langle z,a,b\mid=I$$
where
$$\widetilde{W}_{a,b}(x=|z|^2)=\frac{\N_{a,b}(x)}{\pi}\frac{x^{(b-a)/a}e^{-x^{1/a}}}{a\Gamma(b)}.$$
Instead of shifting the spectrum $e_m$ of the isotonic oscillator backward by $e_0$ if we shift it forward by a positive constant (or if we consider it in its present form) one can also see the resulting CS of type (\ref{s42}) as an analogue of the Mittag-Leffler CS. Let us discuss this point in detail with a more general arbitrary spectrum $x_m=cm+d$ with $c,d>0$. Notice that $c=4$ and $d=2\gamma$ give us $e_m$. In the case of a forward shift of $e_m$ by a constant $e>0$, we have $c=4$ and $d=2\gamma+e$. Let
$$\rho(m)=x_1\dots x_m=c^m(\omega)_m$$
where $\omega=1+d/c$. Thus the states (\ref{s42}) take the form
\begin{equation}\label{s49}
\mid J,\alpha,c,\omega\rangle=\N_{c,\omega}(J)^{-1}\sum_{m=0}^{\infty}\frac{1}{\sqrt{(\omega)_m}}\left({\frac{J}{c}}\right)^{m/2}e^{i(cm+d)\alpha}\mid\psi^{\gamma}_{m}\rangle
\end{equation}
or equivalently
\begin{equation}\label{s410}
\mid J,\alpha,c,\omega\rangle=\widetilde{\N}_{c,\omega}(J)^{-\frac{1}{2}}e^{id\alpha}\sum_{m=0}^{\infty}\frac{z^m}{\sqrt{(\omega)_m}}\mid\psi^{\gamma}_{m}\rangle
\end{equation}
where $z=e^{ic\alpha}\sqrt{{J}/{c}}$. We can observe that (\ref{s410}) is an analogue of (\ref{s48}) with $a=1$ and $b=\omega$, and with the exception that (\ref{s410}) is multiplied by $e^{id\alpha}$. For the states (\ref{s410}), the normalization condition, $\langle J,\alpha,c,\omega\mid J,\alpha,c,\omega\rangle=1$, leads to
\begin{equation}\label{s411}
\widetilde{\N}_{c,\omega}(J)=\sum_{m=0}^{\infty}\frac{|z|^{2m}}{(\omega)_m}=\sum_{m=0}^{\infty}\frac{J^m}{c^m(\omega)_m}=
{}_1F_1(1;w;\frac{J}{c}),
\end{equation}
In this case, the resolution of the identity is justified with the density $\lambda(J)=[\Gamma(1+d/c)c^{1+d/c}]^{-1}e^{-J/c}J^{-d/c}$
using (\ref{s46}). \\
\end{remark}
\section{Conclusion}\nonumber
Wavefunctions of the isotonic oscillator are considered as a basis of a Hilbert space and all classes of CS were constructed as vectors in it. In section 3, two new classes of coherent states were introduced with certain confluent hypergeometric functions and their properties are indicated. Observe that, because of the relations between confluent hypergeometric functions and Hermite polynomials; namely 
$${}_1F_1(-n;\frac{1}{2};x^2)= \frac{(-1)^n n!}{(2n)!}H_{2n}(x),\quad \quad\quad {}_1F_1(-n;\frac{3}{2};x^2)= \frac{(-1)^n n!}{(2n+1)!}H_{2n+1}(x)$$
and the relations between confluent hypergeometric functions and generalized Laguerre polynomials
$${}_1F_1(-n;\gamma+1;z)= \frac{n!}{(\gamma+1)_n}L_n^\gamma(z)$$
our results can be rewritten in terms of Hermite polynomials and generalized Laguerre polynomials with little effort.\\
 A class of temporally stable CS for the isotonic oscillator is presented and the same class is identified as an analogue of the Mittag-Leffler CS \cite{SPS}. CS are discussed with a forward and a backward shift of the energy spectrum of the Hamiltonian. For the backward shift the CS are derived as Gazeau-Klauder CS and for the forward shift the set of CS is identified as a special case of the Mittag-Leffler CS.
\section{acknowledgment}\nonumber
The authors are grateful to referees for their valuable comments. Partial financial support of this work under Grant No. GP249507 from the Natural Sciences and Engineering Research Council (NSERC) of Canada is gratefully acknowledged [NS].

\end{document}